\documentclass[12pt]{article}
\usepackage{a4}
\usepackage{graphicx}
\usepackage{amsfonts}
\usepackage{amscd}
\usepackage{latexsym}
\usepackage{epsf}
\usepackage{amsmath}
\numberwithin{equation}{section}

\newcommand{\be}{\begin{equation}}
\newcommand{\ee}{\end{equation}}
\newcommand{\bea}{\begin{eqnarray}}
\newcommand{\eea}{\end{eqnarray}}
\newcommand{\f}{function }
\newcommand{\fs}{functions }
\newcommand{\eqq}{equation }

\newcommand{\eqqs}{equations }

\newcommand{\nn}{\nonumber}
\newcommand{\fr}{\frac}

\newcommand{\pd}{\partial}

\newcommand{\la}{ \lambda }

\newcommand{\om}{ \omega }

\newcommand{\vp}{\varphi }

\newcommand{\ra}{\rightarrow}

\newcommand{\w}{wave }
\renewcommand{\v}{velocity }
\newcommand{\vs}{velocities }
\newcommand{\wv}{wave velocity }
\newcommand{\wvs}{wave velocities }
\renewcommand{\wp}{wave propagation }
\newcommand{\pv}{propagation velocity }
\renewcommand{\l}{local }
\begin{document}

\title{On a Local Concept of Wave Velocities}
\author{
{\sc I.~V.~Drozdov }\thanks{e-mail: drosdow@uni-koblenz.de} \\
{\small and}\\
A.~A.~Stahlhofen \thanks{e-mail: alfons@uni-koblenz.de}\\
\small  University of Koblenz\\
\small  Department of Physics\\
\small  Universit\"atsstr.1, D-56070 Koblenz, Germany}
\maketitle

\begin{abstract}
 The classical characterization of \wp, as a typical concept for far field phenomena, has been successfully
 applied to many \w phenomena in past decades. The recent reports of superluminal tunnelling
 times and negative group velocities challenged this concept.
  A new local approach for the definition of \wvs avoiding these difficulties while including the
   classical definitions as particular cases is proposed here. This generalisation of the conventional
   non-local approach can be applied to arbitrary \w forms and propagation media. Some applications of the
   formalism are presented and basic properties of the concept are summarized.
\end{abstract}

\section{ Introduction }

The \vs conventionally used for the characterization of \wp are
the phase \v $v_{ph}:=\om/k$, the group \v $v_{gr}:=d\om/dk$, the
front \v $v_{f}:=\lim_{\om\ra\infty}v_{ph}$, the signal \v  $v_s$
with
 $ v_f \ge v_s \ge v_{gr} $, the energy \v $\bar{v}_E := \bar{P}/\rho $ and the phase time \v
 $v_\varphi:= x/ (d\varphi/d\om) $, where $\om$ denotes the angular frequency, $k$ - the wave number,
 $\phi$- a phase shift accumulated in course of propagation, $x$- the distance of propagation, $\bar{P}$ -
 the energy current density and $\rho$ - the energy density.

 This classical concept \cite{brillouin,born} and its improvements \cite{sherman,oughstun} have been successfully applied
  to many \w phenomena in past decades. The recent reports of superluminal and even negative \vs
  \cite{centini, nimtz} the advent of near field optics and the generation of ultra-short pulses \cite{guertler,kohmoto}, for instance,
  revealed intrinsic difficulties and limitations of the classical concept. It is the purpose of the present
  note to resolve these difficulties by a new \l approach for the characterization of \wvs containing the classical
  Ansatz as special case and being applicable to arbitrary \w forms and media of propagation.

   The material is organized as follows: the next section contains a formal discussion of the classical approach
   pinpointing its intrinsic difficulties and summarizing the goals of the new approach. For this purpose
   it is completely sufficient to focus on the notions of phase and group \v. This prepares the ground for the new approach
   outlined in the section 2 and an elucidation of its properties by application to known examples in section 3.

   The behavior of the \wvs  defined here under relativistic transformations is outlined in the section
   4.
   In order to pave the way for applications of the formalism to experimental results like those mentioned above, the discussion is
   concluded by a practical interpretation of the definitions and the extension from a \l  to a global evaluation of signal \vs.

\section{ Some remarks on \wvs }

  The classical concept of \wvs analyses \wp in the far field, where the \w is far away from the source. The
\w is described via propagating harmonic \fs containing a periodic factor of the form $e^{i\vp}$. The argument
 of this periodic \fs is interpreted as a phase and identified with $(t-x/v)$ via the free \w \eqq, where $v$
 is the phase \v \cite{brillouin}.

   A second fundamental concept of characterizing wave propagation
   is that of the "group velocity".
     This concept is - mathematically speaking -
     a comprehensive definition for a specified (linear) superposition of solutions
     of the free wave equation with the same periodicity properties
    usually expressed by the frequency distribution of the constituent
    periodic waves.
     This definition
     is also an inherent far field concept considering "source-free" waves propagating
     in a strongly homogeneous and isotropic medium. This medium is characterized
      only by its "dispersion" (supposed to satisfy additionally the Kramers-Kronig relations),
i.e. by a dependence of the complex index of refraction $n(\om)$ on the periodicity
 parameter of the wave - the frequency $\om$.

  The different notions of \vs (and the accompanying notion of a signal) developed in this context were a major step
  forward allowing to accomodate many experimental data and to characterize many \w phenomena. (cf. e.g. \cite{born})
   It is, however, not obvious to what extent these far field concepts can be applied to near field problems or any
   of the other challenges mentioned above. A possible answer to this question simultaneously preserving the classical
   concepts whenever applicable is the focus of the present work. The basic idea is to employ a \l definition of a
   \pv.

Such local approaches can be found in many areas physics and its
applications in concerned natural sciences.
 A need for an alternative local approach to the analysis of \w propagation was justified already
  since years in different fields of application such as geophysics \cite{aric}, physical chemistry
   \cite{hoffman,rabadi}, biophysics \cite{xioaoming}, acoustics \cite{simmons},
    dealing with a relative "slow" waves, compared to the optics and electronics, thus being long
  ago in the near-field region, where the actual problems require a fine resolution of wave behavior.

There is thus a justified reason for a \l approach to \wvs since
this allows to resolve the intrinsic problems accompanying the
classical global definitions. To elucidate this remark we now list
some of these problems.

 The conventional analysis of \wp is based  on a representation via periodic \fs by means of Fourier analysis. A well
 defined phase \v is assigned to each Fourier component equipped with its frequency;
 a corresponding group \v is subsequently defined for the complete Fourier superposition, i.e. for a \w packet.
  The classical \w theory is based on such definitions relying canonically of the phase of periodic \fs and on group
   of such \fs.
  This Ansatz is supposed to be the inherent kernel of any characterization of \w phenomena
  and has also been used as a guideline for characterizing general
  - not necessary periodic - electromagnetic pulses \cite{brillouin}.  A close inspection, however, reveals
   several shortcomings of this Ansatz with respect to mathematics and
  physics.

  Let us start with the local features of \wp. A propagating wave is per definition a \l space-time distribution
  which is itself a solution of a \l differential \eqq. A definition of a \wv as a space-time relation should respect
   this frame being \l as well. This is not the case in the conventional definitions of \wvs: these involve the frequency
    and the \w vector, which are (temporally and spatially) nonlocal parameters in following the independent space- and
    time-periodicity. Since these parameters do not enter the \w \eqq, they are usually plugged in from the outside. The source
    free \w \eqq for instance, admits solutions of arbitrary shapes without an {\it a priori} assumption of periodicity.

     Simultaneously, the classical approach always takes recourse to a representation of an arbitrary solution by a set of
     periodic harmonic \fs, where each element of the set does in general not obey the \w \eqq. Any attempt to apply this
      strategy to waves resulting from non-linear \eqqs such as solitons, for instance, elucidates this problem  in a clear
      way since in this case even a linear superposition of solutions is not anymore a solution.

Second,
  the subject of transmission and velocity of a signal is based on sufficiently local procedures of measurement \cite{guertler}.
  Roughly expressed, the intervals between several space-time points are measured.
 Definitions of \vs based on periodicity parameters can possess generically neither time- nor space-locality.
 Moreover, any Fourier transformation is basically a global object, and all manipulations concerned are
 in general mathematically exact only with integration over the whole space and whole time, as well as
 over an infinite frequency band.

   The conventional definitions of the phase and group velocity mentioned above \cite{brillouin} are based on
   the special case of a propagation of periodic harmonic waves in homogeneous media. Thus there is already an essential
    contradiction between the local character of wave propagation and the inherent globality contained in the definitions of
    wave velocities.

Any attempt to construct a local measurable object using Fourier sums or integrals contain an essential contradiction
as outlined above and provides indeed no real locality.
  This is the source of several problems arising when replacing originally local features by
 "microglobal" ones \cite{sherman, kohmoto2002, bloch}.

 Even if the approach is supported by mathematical consistency
 (like the assumption of an infinite frequency band \cite{brillouin}),
  it does not lend itself easily to a transparent physical interpretation.
  A nice example thereof is the phenomenon of signal transmission in (optically active) media,
 which triggered a lot of discussions \cite{sherman, loudon, shiren}

 The classical papers \cite{brillouin} and later improvements thereof \cite{bloch, oughstun}
  still contain an essential mismatch between local and global wave features, based on several mathematically correct
  but physically misleading definitions
   (like the notion of front and signal \vs \cite{bers, pacher}).
  This Ansatz is bound to lead to controversial results in applications. For example, it seems not more to be surprising,
  that the subject  of "group velocity" fails to describe a propagation of ultra-short pulses \cite{kohmoto}.\\

These shortcomings of the classical concept of \wvs motivated the
Ansatz presented here for the following reasons:

- The classical harmonic approach represents local wave features
in terms of generically non-local attributes
\cite{sonnenschein,schriemer}. Hence:

- The mathematical equipment is not exact for finite physical
values (like a limited frequency band);

- It is therefore not obvious how to apply it to near-field
effects, if the size of space-time regions
 are of an order of magnitude smaller than the wave periodicity parameter;

- There is no canonical application thereof for inhomogeneous and
anisotropic media, as well as for ultra-short pulses of arbitrary
form and for nonlinear waves \cite{kohmoto2002, gomez};

- As a consequence, the applicability of this approach for several
fields of modern quantum optics, nano-optics
 and photonics is difficult, since these topics deal with parameter areas outside its range of
 validity \cite{centini}. Further motivations outside of optics have been mentioned above.

 The classical approach is also to some extent problematic
  from a pure technical point of view:
  it is an essential restriction of generality and universality of the
 theory enforcing via the Fourier analysis - a number of
 problematical artifacts. A prominent example thereof being the assumption of infinite frequency bands for signals.
 These additional problems are then resolved via the analysis of dispersion relations.

 A first step towards a more general concept should be based on an independent alternative
  approach without doubting the canonical harmonic criteria, i.e. leading to the same (verified)
  results in known cases.

 The aim of the present paper is to establish a transparent criterion for evaluation
  of the propagation velocity for a quite arbitrary signal, that does not need an explicit
   representation in terms of periodic or exponential functions at all, and with minimal loss of
   generality in other respects.
 First of all, we have to recall, what is being measured in experiments and what is meant
 when speaking about a "wave velocity", thereby providing the spectrum of wave velocities.

\section{ N-th order phase velocity (PV)}
\label{ PVs}

  The following discussion is restricted to a 1+1 dimensional space-time for simplicity.
    To evaluate the speed of a moving matter point one has to check the change of the
    space coordinate $\Delta x$ during the time interval $\Delta t$. How can this Ansatz be applied
    to waves ?

  An arbitrary wave is a \f defined on the 2-dimensional space-time $(x,t)$ and
  one has no {\it a priori } defined fixed points to follow up as in the former case.
  Let us therefore analyze a continuous smooth \f $\psi(x,t)$ of arbitrary shape.\\

   In a first step towards a local definition of velocity we consider a vicinity ${\cal U}(x)$ of a
    certain point $x$ at the certain time $t$. Further we assume the local information
    about the \f $\psi(x,t)$ to be measurable, i.e. we should be able (at least in principle) to evaluate
    the values of the \f $\psi(x,t)$ itself and all its derivatives in the point $x$ as well,
 and in the vicinity  ${\cal U}(x)$.

  A description of propagation is based on
  monitoring the fate of a certain attribute ("labelled point") of this shape, i.e.
   one finds the same attribute at the next moment $t_1$
  in the next point $x_1$.

 \begin{figure}[h]
\epsfxsize=10cm\epsffile{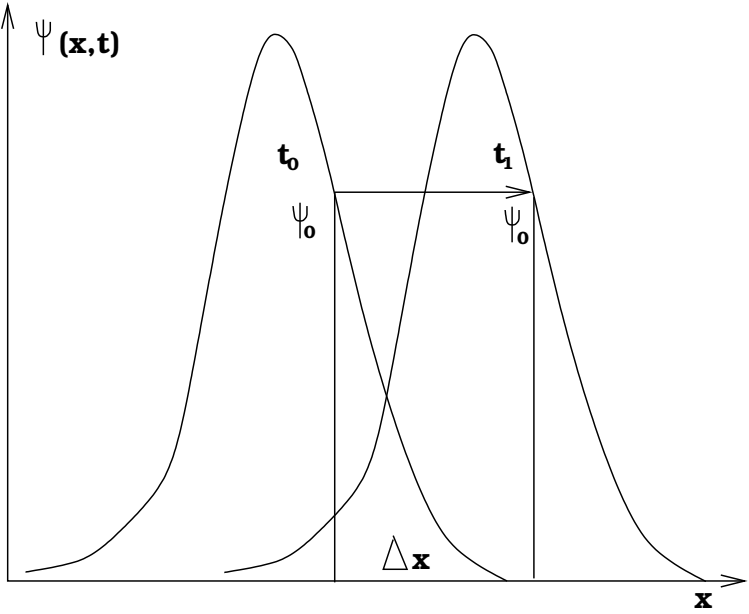}
\caption{\small The propagation of a signal $\psi(x,t)$ without a change of shape with the value $\psi_0$ of
 the amplitude as an attribute to be traced in course of propagation}
\label{normalshape}
\end{figure}

 Let this
   attribute be labelled by some one-point fixed value $\psi_0$ of the \f $\psi(x,t)$.
 Suppose, one follows up this local "attribute" and manages a local observation of the condition:

 \be
 \psi(x,t)-\psi_0=0,
 \label{implicit}
 \ee

 which fixes the space-time points $\{ x, t\}$ where this
 condition holds (Fig.\ref{normalshape}). Thus we have a \f $x(t)$ given in an implicit form in eq.(\ref{implicit}).

  The first order implicit derivation provides the velocity of this one-point local attribute:
 \be
 v_{(0)}(x,t):=\fr{dx}{dt}= -\fr{\fr{\pd\psi }{\pd t}}  {\fr{\pd\psi }{\pd x} },
 \label{0ord}
 \ee
called from here on the "zero-order" or "one-point" phase velocity (0-PV).

 (Here the subject "phase" is treated in sence of local charactreristics in the point of shape
 including all N-order derivatives, $N=0,1,2,...$ i.e. all terms of the Taylor expansion in this point).

 It describes in the simplest case the speed of translation of some arbitrary
 pulse, that can  be treated as the signal propagation velocity, provided that the measured
  value $\psi_0$ of the amplitude is the signal considered.\\

 To proceed with the local description of the shape propagation we now consider
as
  the measurable attribute of $\psi(x,t)$ at some time $t_0$ not the
single value $\psi_0=\psi(x_0;t_0)$ at the point $x_0$, but a set of values $\psi$ in a local
 neighborhood (like ${\cal U}(x)$) close to this point.
  Then another local attribute can be constructed to trace their propagation.
 If, for instance, one looks at a certain value of the first derivative $\pd\psi(x; t_0)/ \pd x:=\chi_0$ of the
 shape $\psi$, (like $\pd\psi(x; t_0)/ \pd x=0$, i.e. at a maximum or minimum point), one should
  consider the propagation of the condition:
\be
\fr{\pd\psi(x,t)}{\pd x}=\chi_0;
\ee
this provides its propagation velocity via
\be
v_{(I)}(x,t):= -\fr{\fr{\pd^2\psi }{\pd t\pd x}}  {\fr{\pd^2\psi }{\pd x^2} },
\label{1ord}
\ee
called in view of (\ref{0ord}) the "first order" or "two-point" phase velocity (1-PV) respectively.

When tracking the propagation of a maximum (minimum), the conditions $ \fr{\pd\psi(x,t)}{\pd x}=0 $  and
$\fr{\pd^2\psi }{\pd x^2}<0$ have to hold simultaneously.

 This Ansatz is easily iterated to phase velocities of order $N$ interpreted as
 the propagation velocities of higher order local shape attributes via


 \be
 v_{(N)}(x,t):= -\fr{\fr{\pd^{N+1}\psi }{\pd t\pd x^N}}  {\fr{\pd^{N+1}\psi }{\pd x^{N+1}} },
 \label{Nord}
 \ee
leading to N-th order ("N+1-point") PV, describing the propagation of higher order local shape attributes.

  The phase velocities so defined are obviously local features depending on space-time
 coordinates. It has to be noted, that any given problem at hand might require a particular choice of a PV
  allowed for by the definitions given above. The following examples elucidate these requirements and demonstrate
  the flexibility of these definitions.

  Before proceeding to it should be recalled, that the PV-spectrum
  has been obtained wanting to be able to describe the propagation of an arbitrary pulse in terms of
 local attributes in a medium whose properties depend on several variables, especially time and space
  coordinates being the most prominent and natural examples thereof.
   Any initial shape $\psi(x_0,t_0)$ thus should be deformed during the propagation (or evolution, as typicaly
   encountered for dispersive media). The ordinary phase velocity $v_{0}$ therefore is not a relevant
  criterion to characterize the shape propagation and one has to choose an appropriate PV $v_{(N)}$.

\begin{figure}[h]
\epsfxsize=10cm\epsffile{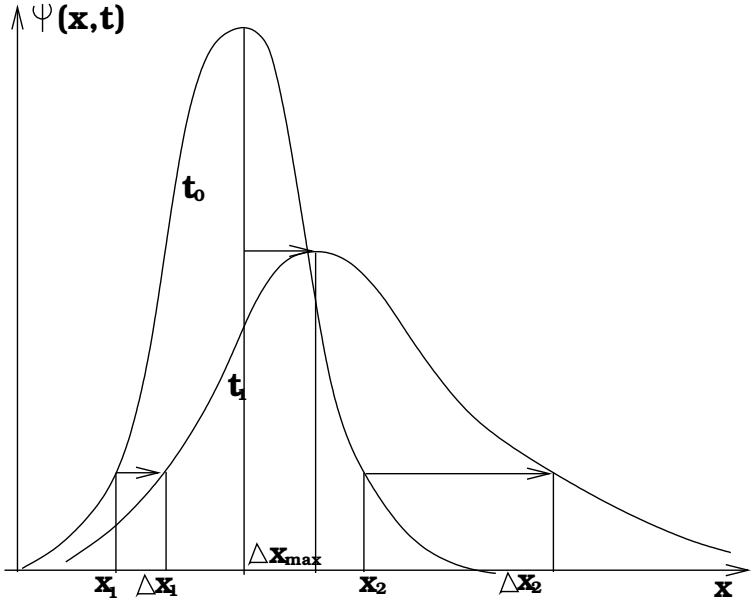}
\caption{\small A damping deformation of a signal $\psi(x,t)$. A natural attribute allowing to describe propagation
is the peak location (maximum)}
\label{damped}
\end{figure}

   For example, let the pulse $\psi(x,t_0)$ be subject to damping (Fig.\ref{damped}).
  As a concequence, the zeroth order PV measured in the point $x_1$ gives a magnitude much smaller
  as the same magnitude measured in the point $x_2$.

\begin{figure}[h]
\epsfxsize=10cm\epsffile{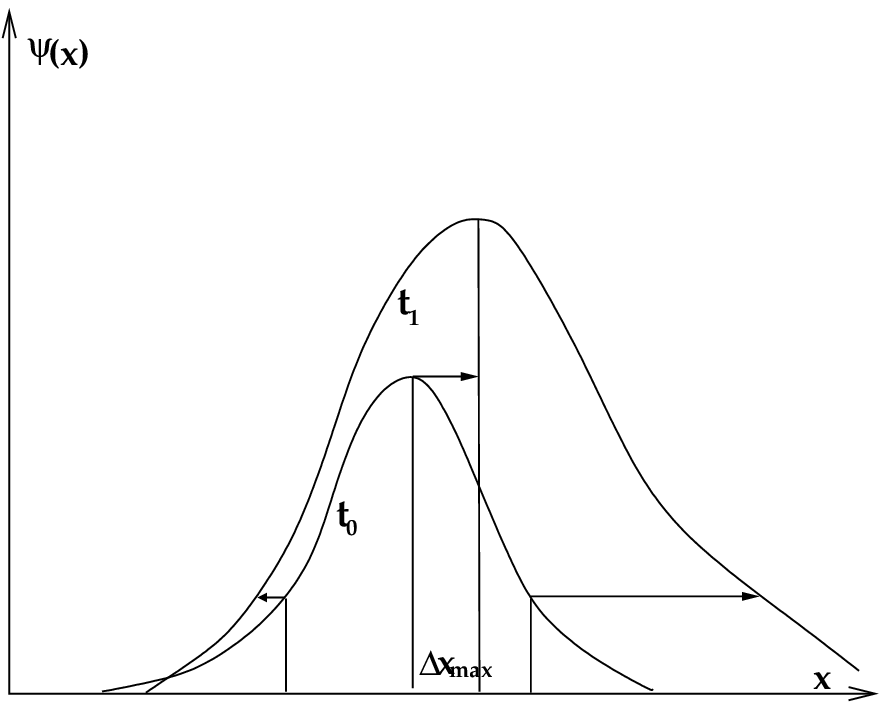}
\caption{\small The propagation of an amplified signal $\psi(x,t)$. The location of the maximum describes the wave propagation}
\label{excited}
\end{figure}

   For an amplified signal (Fig.\ref{excited}, like a signal propagating in a laser excited medium), by comparison, the
   zero order PV from the point $x_1$ provides altogether even a backward propagation.
   In both cases an appropriate approach would be to apply the first order PV
   $v_{(I)}$ which describes the propagation of the maximum up from the point $x_0$
   properly.\\

\begin{figure}[h]
\epsfxsize=10cm\epsffile{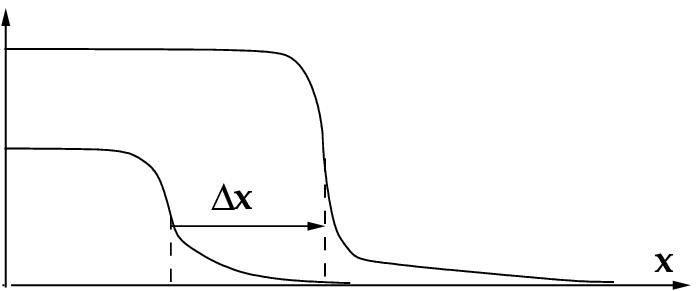}
\caption{\small The propagation of a growing kink ("tsunami model").
 The turning point is chosen to trace the propagation }
\label{kink}
\end{figure}

     For a propagation of a kink front that experiences a deformation, one
     can check the translation of the second derivative of the shape, keeping
      track of the turning-point of the kink shape (Fig.\ref{kink}). In this case the second order ("three-point")
      PV turns out to be the relevant velocity of propagation.\\

Summarized, the local concept is based on local attributes of the \w shape being traced in course
of propagation. The phase of an arbitrary \w in some point is characterized by the local Taylor expansion
in this point. The local attribute of N-th order is therefore the N-th spatial derivative of the \w shape.
 The phase velocity of N-th order (N-PV) is the \v of propagation of the corresponding N-th order attribute.


 Finally, it should be noted, that the definiton of phase velocities of order zero and one,
 the  $ v_{(0)}$ and $v_{(I)}$ respectively, admits a straightforward generalization to two-,
 three-, and higher-dimensional propagation, while the phase velocities of second ( $ v_{(II)} $)
 and higher orders
 inherently contain a certain element of ambiguity
 in their definition, since a possibility to choose
  a second-order attribute to be traced is not unique is not unique \cite{dimension}.


\section{ Examples}
\label{ samples}

 Let us consider the conventional 1+1-dimensional wave equation
   \be  \left[ \fr{1}{a^2}\fr{\pd^2}{\pd t^2}-\fr{\pd^2}{\pd x^2} \right]\psi(x,t)=0
   \label{free-eq}
   \ee

    possessing translational solutions of the form
    \be
    \psi(x,t)=\psi( t \pm \fr{ x}{a} ).
    \label{transl-mode}
    \ee
  It is easy to check that in this case the PV's of all orders defined above are
  identical and read

  \be
   v_{(N)}(x,t)= a ,\ \ \ \ N=0,1,2,...
  \ee

  which is nothing else but the classical phase velocity
  $\pm a$, thereby satisfying the {\it a priori} definition
   of "phase velocity" itself as a medium constant in (\ref{free-eq}).

 Let a propagating shape $\Psi$ now be subject to a temporal damping similar to (Fig.\ref{damped}) with
 \be
 \Psi(x,t)=\psi(t-\fr{x}{a})e^{-\la t}\equiv \psi(\phi)e^{-\la t}
 \label{t-damped}
 \ee
that obeys the wave \eqq (a typical field \eqq with a dynamical dissipation and a mass term)
\be
\left\{ \fr{1}{a^2} \fr{\pd^2}{\pd t^2}-\fr{\pd^2}{\pd x^2}+
 \fr{2\la}{a^2}\fr{\pd}{\pd
 t}+\fr{\la^2}{a^2}   \right\}  \Psi(x,t)=0
  \label{damped_eq}
\ee

 The ordinary 0-PV velocity reads

 \be
 v_{(0)}=a\left( 1-\la\fr{\psi}{\psi'} \right),
 \ee

 where the prime denotes the derivative of $\psi$ with respect to its argument $\phi \equiv t- x/a $.

  This result provides a velocity with bad physical features: the velocity grows for a descending shape, for an ascending
  shape it decreases, can even be
  negative, and it diverges exactly for the peak point (under the condition that the (measured) amplitude $\Psi$ as well as
   parameter $a,\la $ have positive values).

 A relevant physical velocity in this case is for instance the 1-PV
 \be
  v_{(I)}=a\left( 1-\la\fr{\psi'}{\psi''} \right)
 \ee
providing for a peak being traced exactly the canonical phase velocity entering in the wave \eqq (\ref{damped_eq}).

For this shape further PV's of higher orders are given by (\ref{Nord}):
 \be
  v_{(N)}=a \left(1-\la \fr{\psi^{(n)}} {\psi^{(n+1)}} \right) \equiv a \left(1-\la (\log' {\psi^{(n)} })^{-1} \right)
 \ee
where $\psi^{(n)} $ denotes the $n$-th derivative of $\psi$ with respect to its argument as mentioned above.
For an amplified signal as in Fig.\ref{excited} we can e.g. change the sign of $\la$.
 Especially, for the case of a kink (Fig. \ref{kink})

 \be
 \psi( t - \fr{x}{a} )\equiv \psi(\phi)=\arctan \phi,
 \ee
the spectrum of phase velocities reads by comparison :

 \bea
&& v_{(0)}=a\left( 1 + \la (1+\phi^2) \arctan \phi  \right) , \nn\\
&& v_{(I)}= a\left( 1 - \la \fr{1+\phi^2}{2\phi}  \right) ,\\
&& v_{(II)}= a\left( 1 - \la  \fr{\phi^3+\phi}{3\phi^2-1}  \right) .\nn
 \eea

 The ordinary 0-PV has an oscillating sign at  $ \la  $ and is multiple defined because of the {\bf\it arctan} \f.
 Therefore it cannot be interpreted as a well-defined physical velocity.
 If the point being traced should be labelled by a derivative attribute,
  it turns out to be a physically inconvenient choice
 since the shape possesses no real peaks, whose propagation could be traced.
  Moreover the velocity $v_{(I)}$ diverges at $\phi=0$.

 The possible labelled attribute is also the turning-point traced by the 2-PV.
  The velocity $ v_{(II)}$  also possesses two singularities at $\phi=\pm 1/\sqrt{3}$ which do not
 coincide with the labelled point $\phi=0$, so it can be successfully followed up at the measurement.\\

 {\it Historical remark \\}

  The definition of a \wv
  proceeded with an extension of a translational
   solution of (\ref{free-eq}):

   \be  \psi(x,t)=\psi( t - \fr{ x}{a} ) \equiv \psi[ \fr{1}{\om} ( \om t -\fr{\om}{a} x ) ] ,\equiv \phi( \om t -k x ),\ k\equiv
   \fr{\om}{a},
   \ee
  where the parameter $\om$ has been interpreted as a frequency of a necessary perodic harmonic \f $\phi$.
  usually $e^{\pm ix}$ as mentioned above in the introduction. It is not surprising that the 0-order PV provides in this case
  the value $\om/k$, canonically interpreted as a phase velocity of periodic wave

 In case of any interrelations between $k$ and $\om$
 that are not encountered in the wave equation, in particular any dispersion relation
  between frequency and wave-number,
 the PV $v_{(0)}$ is in fact a proportionality factor
  \be
d\om= v_{(0)}dk,
  \ee
which is identical with the classical definiton of a group velocity \cite{brillouin}.

  At this point it should be recalled, that the original idea of the group velocity $U$, as summarized e.g. in \cite{bateman},
  \be \fr{\pd \la}{\pd t}  + U\fr{\pd \la}{\pd x} =0  \label{group} \ee
  was of the
  similar form as the recent definition (\ref{0ord}) of 0-PV.
It should be noted here that \eqq (\ref{group}) involves local derivatives of the non-local parameter,i.e. the wavelength $\la$.
    For a variable wavelength much smaller than a vicinity of the point, this group velocity provides therefore a
    natural approximation to the zero-order phase velocity $v_{(0)}$.

  In the present approach
  this feature appears as a physical phase velocity following in a straightforward way from the
  interpretation of phase propagation and does not require an
  interpretation of $\om$ and $k$ as a frequency and wave number, as well as a
   constancy of some group respectively.

  It is noted in passing, that
  in the case of a kink there are no suitable definitions of a wave group and of a concerned group velocity for this solution
  since a Fourier decomposition does not work on a non-compact support.

 \section{  Lorentz covariance }
 \label{ lorentz}

  The zero order PV (\ref{0ord} ) possesses an interesting
  property, i.e. a local covariance in the sense of special relativity,
  as shown now. This is not the case for PV's
  of higher orders.

   The PV $v_{(0)}$, measured in some stationary system $X$, takes in some other system $X'$,
    moving with a constant speed $V$, via the Lorentz transformations
   \be
   x'=\fr{x-Vt}{\sqrt{1-\fr{V^2}{c^2}}},\ \  t'=\fr{t-\fr{Vx}{c^2}}{\sqrt{1-\fr{V^2}{c^2}}}
   \ee
the form
  \be
  v'_{(0)}=\fr{v_{(0)}+V}{1+\fr{v_{(0)} V}{c^2}}.
  \ee

 This means that the zero order PV respects the relativistic velocity addition.
 Especially, a subluminal zero order PV remains also subluminal in any other
  moving system $X'$.

  This result should not be a surprise, since the definition of the $ v_{(0)}$
 is implied by the condition:
 \be
  \psi(x,t)=const,\ \
  \ee
  which remains to be of the same form under arbitrary transormations\\ $x=x(x', t');\  t=t(x', t')$, implying
  \be
   d\psi(x,t)\equiv \fr{\pd\psi }{\pd t}dt+\fr{\pd\psi }{\pd x}dx=0
   \label{differential_0}
 \ee
for the first order differential form (or simply first differential), which possesses a form-invariance property under
 transformations.

 Since \be v_{(0)}(x,t)=\fr{dx}{dt}\ee per definition, the \eqq (\ref{differential_0}) inplies the definition (\ref{0ord}) of 0-PV,
  so it should behave under space-time transformations as a usual velocity of a matter point.

 The first order PV (\ref{1ord}), by comparison, evaluated in some system $X$
  transformes in the moving system $X'$ to:
  \be
  v'_{(I)}= -\fr{\left\{\left( 1+\fr{V^2}{c^2} \right) \fr{\pd^2}{\pd x\pd t} - V\left(  \fr{1}{c^2}\fr{\pd^2}{\pd t^2}+ \fr{\pd^2}{\pd x^2} \right)\right\}\psi }
               {\left\{\fr{V^2}{c^4}\fr{\pd^2}{\pd t^2}  + \fr{\pd^2}{\pd x^2} -2  \fr{V}{c^2}\fr{\pd^2}{\pd t\pd x}\right\} \psi }.
  \ee

If the pulse $\psi (x,t)$ obeys the free wave \eqq  (\ref{free-eq}), the transformation becomes

\be
 v'_{(I)}=-\fr{ \left(1+\fr{V^2}{c^2} \right)v_{(I)}+2V }{\left(1+\fr{V^2}{c^2} \right)+\fr{2V v_{(I)}} {c^2} },
\ee
i.e. a relation that should be called the "first order velocity addition". It differs obviously from the corresponding transformation
 of $v_{(0)}$, since the definition of the velocity $v_{(I)}$
 results from the condition

 \be
 d\left( \fr{\pd\psi (x,t)}{\pd x} \right)=0
 \ee
 whose form is explicitely non-invariant under space-time transformations.

 It can be shown that a subluminal first order PV is still preserved by this transformation as well. Note, that for a signal which does not obey the
 free \eqq (\ref{free-eq}), this restriction is in general not guaranteed anymore.\\

\section{  A global velocity of signal transmission and "dynamic separation"  }
\label{ global}

  The discussion of local velocities was aimed towards an evaluation of global
  features of signal propagation, namely the propagation through a finite spatial interval
   during a finite temporal interval.
  In other words, a global velocity, practically measured, means roughly the length of
 the  distance $\Delta x$ traveled by a traced attribute divided by the time interval $\Delta t$.

\begin{figure}[h]
\epsfxsize=10cm\epsffile{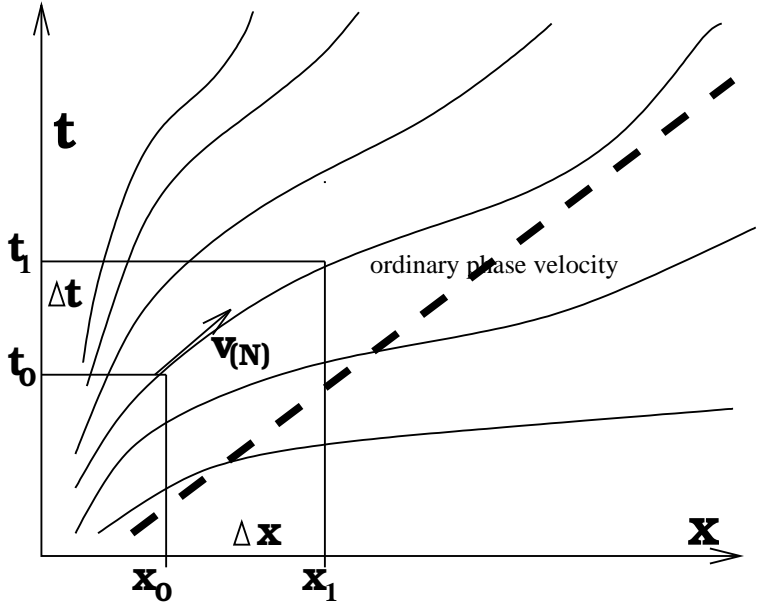}
\caption{\small Phase velocities as a family of isoclines $v_{(N)}(x,t)$ and averaged global velocities between two
events (measurements)}
\label{isoclines}
\end{figure}

 The local PV's of N-th order analyzed above can be interpreted as a first order differential
 \eqqs of the form
  \be
   v_{(N)}(x,t)=\fr{dx}{dt},
   \label{def-vN}
   \ee
that can be illustrated graphically as a field of {isoclines} (Fig.\ref{isoclines}).
Here the local PV is the tangent function of the tangent vector of the isocline, and the averaged (total) velocity
between $(t_0,x_0)$ and  $(t_1,x_1)$ is represented by the tangent of the
hypotenuse of the triangle $\{ (t_0,x_0),(t_1,x_1), (t_0,x_1)\}$, (cf. Fig.\ref{isoclines}).

  This procedure is elucidated best
  by some clear and well known examples. Consider the propagation of a translation mode of the form

   \be
     \psi(x,t)=\psi(\xi t - k(x)) :=\psi(\phi)
     \label{transl}
   \ee
describing, for instance, an electromagnetic field propagation in an inhomogeneous dielectric medium.
 In this case, if the transversality of the \w not assured, the differential \eqq for any components of the
 field can possess e.g. a form like this:
 \be
 \fr{n^2(x)}{c^2} \fr{\pd^2}{\pd t^2} \psi-\fr{\pd^2}{\pd x^2} \psi+\ln'n(x) \fr{\pd}{\pd x}\psi=0
 \ee
where $n(x)$  is the spatially dependent index of refraction. From the Ansatz (\ref{transl}) it results for the $k(x)$
from the optical inhomogeneity over the spatial coordinate $x$:
   \be
   n(x)=\fr{c}{\xi}k'(x),\ \ \ \  k(x)= \fr{\xi}{c} \int n(x)dx,
   \ee
Then the local zero order PV is provided by eq.(\ref{0ord}) and results in the first order \eqq
    \be
    v_{(0)}(x,t)\equiv\fr{dx}{dt}= \fr{c}{n(x)},
    \label{0PV_local}
    \ee
that is a quite trivial result.

 We proceed now with phase velocities of higher orders.
For the 1-PV velocity $ v_{(I)}$ defined by eq.(\ref{1ord}) the same procedure leads to
     \be
      v_{(I)}(x,t)\equiv \fr{dx}{dt} = \fr{c}{ n(x)-\fr{c}{\xi}\fr{\psi'}{\psi''} \ln'n(x) }
     \label{localDD}
     \ee
where $x$ and $t$ denote the space and time location of the first oder attribute being traced;
 $\psi'$and$\psi''$ are the first and the second derivative of the \w \f $\psi$  with respect to its
 argument $\phi$ (cf.\ref{transl}).
Similarly, for the 2-PV we obtain:
 \be
  v_{(II)}(x,t) = \fr{dx}{dt} = \fr{c}{ n(x)-3 \fr{c}{\xi}\fr{\psi''}{\psi'''} \ln'n(x) +  \fr{c^2}{\xi^2}\fr{n''}{n^2} \fr{\psi'}{\psi'''}}.
     \label{localDDD}
 \ee

 Suppose, we measured at the moment $t_0$ some value of the fisrt order attribute $\psi'(x_0)$ in the point $x_0$,
  and we are interested, where is this attribute to recover at the next moment $t_1$. For a maximum or a minimum point
 the result does not distinguish from the zero order velocity, as in the case considered above. Now, let we
  measured at $t_0$ some values:

    \be \psi'(0,x_0):=\psi_0' \mbox{ and  } \psi''(0,x_0):=\psi_0'';\ \ \ \Lambda:=\psi'(0,x_0)/ \psi_0'' \ee
An explicit integration of eq.(\ref{localDD}) leads to
   \be
    c(t_1-t_0)= \int\limits_{x_0}^{x_1} n(x)dx-\Lambda\fr{c}{\xi}\ln\fr{n(x_1)}{n(x_0)}
    \label{globalDD}
   \ee

for the interval of two events, where the signal attribute choosen for tracing
    has been checked at the time $t_0$ in the point $x_0$ and afterwards in $x_1$ at
   $t_1$.
     Let us assume that the point $(t_0,x_0)$ was not a critical point (i.e. of the knot type)
 of the \eqq (\ref{def-vN}). Then the global transition velocity
 between two points $x_0,  x_1$ is evaluated unique in such a way: divided by the spatial interval $x_1-x_0$ the eq.(\ref{globalDD} )
 provides the factor $\gamma$:
  \be
 \gamma:=  c\fr{t_1-t_0}{x_1-x_0}= \fr{1}{x_1-x_0} \left[\int\limits_{x_0}^{x_1} n(x)dx-\Lambda\fr{c}{\xi}\ln\fr{n(x_1)}{n(x_0)}\right]
  \ee
It follows, that a value of $\gamma$ less than 1 is not prohibited
in principle even for real $n>1$, that means a superluminal
averaged global velocity along the finite distance. A similar
result appears for 2-PV and higher orders. A numerical examples
for dielectric media of special modulation confirms this
possibility \cite{dielectric}. The discussion of physical
interpretation of this result is still open.

It is also worth to notice, that for the first order PV (as well as the second and higher orders) in media with
  a variable refraction index $n(x)$ an essential dependence on the parameter $\xi$ enters.

 The meaning of $\xi$ can be derived from the given form of the solution $\psi$.
It can be interpreted, for instance, as a frequency factor for a periodic mode or a damping factor for
an evanescent mode.
    This parameter simply means an enumeration of solutions of a solution family (space of solutions).
The merit gained above is the separation of these solutions with respect to the parameter $\xi$ by the
different first (and higher)-order PV's through the inhomogeneity of the medium.

  In the case of a periodic wave for instance, it can be interpreted as a dispersion
    although an explicit frequency-dependence of $n(x)$  was not assumed.
    Moreover, the parameter $\xi$ has not necessarily to be interpreted as a frequency of some
    temporally periodic oscillation, but rather as a time component of the space-time wave vector $\{\om, {\bf k}\}$
    for a translation mode (\ref{transl-mode}) in a general form. The phenomenon considered thus
    has another origin and is much more general as a conventional non-localized frequency dispersion
    $n(\om)$

     Thus we established for the first (and higher) order PV's in optical
     inhomogeneous media the essential dependence on the t-component $\om$
      of the wave translation vector, especially frequency, even for local non dispersive
     media, which could be called "dynamic separation". The global averaged dynamic
    separation of eq. (\ref{globalDD}) survives as a corollary of the local separation of
     (\ref{localDD}). This phenomena does not occur for the ordinary zero order
     PV.

 \section{Concluding remarks }

  The inherent inconsistency of the classical subjects of phase velocity and group velocity,
  (as well as signal velocity based therein) has been discussed.
   The inapplicability of these concepts for actual studies in photonics, near-field
   and nano-optics has been shown to result from the essential non-locality of these definitions.

 An alternative approach for description of a propagation velocity has been proposed.
 It is strictly local and is based on the natural assumption of an ordinary measurement procedure.
  It does not need any {\it a priori} condition
 of periodicity, frequency, groups and packets or other canonical attributes.

 The definitions presented above are applicable in a natural way for arbitrary waves and pulses.
  In a mathematical sense they describe a propagation of a perturbation in any field in space-time.
   Examples could be: an acoustic wave as a pressure perturbation,
 a gravitation wave on a fluid surface, a spin wave in a solid state a.s.f.
 In particular,
 the formalism is very suited for the description of
 particle propagation in field theory, where particles are considered as field
 perturbations.

  This approach results in the set of measurable propagation velocities.
 In zeroth order the propagation velocity coincides with the ordinary phase velocity and appears
 to be ordinary Lorentz covariant; further application
 gives rise to generate "Lorentz covariance of higher orders".

  For propagation in inhomogeneous media, for instance for light in a medium with a space-dependent
   index of refraction, phase velocities of higher orders show some remarkable properties:

   a) they are not restricted in general by the speed of light in vacuum;

   b) they exhibit an essential dependence on the time component of the wave vector solely as a result
    of inhomogeneity, treated
    canonically as a "dispersion". This appears to be a more general phenomena,
    and does not presuppose any periodic frequency and dispersive properties of media.
    It should be called for this reason "dynamic separation" \cite{dielectric}.


 Further applications, like photonics and optoelectronics are
 dealing with signals i.e. non-localized objects, represented as a cut of wave shape between labelled points
  possessing a certain spatial length (or temporal duration).
  These two points are marked by a value of zeroth order attribute (e.g. a voltage amplitude), or by a first order one (a block between
  an ascent and a descent of amplitude).
  In this sense the signal is considered as
  not a local, but at least a bi-local object. Its propagation can be traced by an averaged bilocal velocity,
   constructed from local ones in a natural way.

 The same concerns a propagation of single photons, especially of low energy, which are non-localized objects.
  For the case of evanescent modes we are dealing yet with a sufficient delocalization.

 Another approach to the signal transmission is based on a pure local signals, which can be, for instance, a point of discontinuity,
  propagating away from the perturbation \cite{stenner}. For these problems the \vs of higher orders, especially 1-PV and 2-PV provide
  a suitable tool, as shown above. In this context are the velocities 1-PV and 2-PV of special interest, being treated in pure
  physical sense as a true speed of interaction.


\end{document}